\documentclass[journal,final]{IEEEtran}
\usepackage{amssymb}
\usepackage{amsmath}
\usepackage{graphicx}
\usepackage{bm}
\usepackage{amsfonts}
\usepackage{cite}

\def\be{\begin{equation}} 
\def\ee{\end{equation}}

\begin{document}

\title{Investigating surface loss effects in superconducting transmon qubits}

\author{J. M. Gambetta,~\IEEEmembership{Member,~IEEE,}
	C. E. Murray,
	Y.-K.-K. Fung,
	D. T. McClure,
	O. Dial,~\IEEEmembership{Member,~IEEE,}
	W. Shanks,
	J. Sleight,~\IEEEmembership{Senior Member,~IEEE,}
	M. Steffen,~\IEEEmembership{Senior Member,~IEEE}
\thanks{J. Gambetta, C. Murray, Y.-K.-K. Fung, D.T. McClure, O. Dial, W. Shanks,
J. Sleight and M. Steffen are with International Business Machines, IBM TJ
Watson Research Center, Yorktown Heights, NY, 10598 USA e-mail: (jay.gambetta@us.ibm.com).}
}

\maketitle

\begin{abstract}
Superconducting qubits are sensitive to a variety of loss mechanisms including
dielectric loss from interfaces. By changing the physical footprint of the qubit
it is possible to modulate sensitivity to surface loss. Here we show a
systematic study of planar superconducting transmons of differing physical
footprints to optimize the qubit design for maximum coherence. We find that
qubits with small footprints are limited by surface loss and that qubits with
large footprints are limited by other loss mechanisms which are currently not
understood.
\end{abstract}

\IEEEpeerreviewmaketitle

\section{Introduction}

\IEEEPARstart{I}{mproving} coherence times (or quality factors) of
superconducting qubits is necessary in order to implement
meaningful tests of quantum information processing in such systems. It has been
known for many years that dielectric loss can play a potentially limiting role
for qubit coherence \cite{Martinis05}. While other loss mechanisms such as
quasi-particle loss \cite{Sun12,Wang14}, limitations due to infra-red radiation
\cite{Barends11,Corcoles11}, losses related to the qubit readout \cite{Reed10a},
or electromagnetic radiation \cite{Sandberg13} can play dominant roles, a
picture has been emerging recently that dielectric loss at surfaces also plays a
key role \cite{Wenner11,Geerlings12,Megrant12,Quintana14}. In particular it has
been observed that quality factors tend to increase when the physical footprint
of the devices increases for both resonators \cite{Geerlings12,Megrant12} and 3D
qubits \cite{Dial16}. This points to surface loss as a detractor of qubit
coherence, suggesting that qubits can be optimized for coherence by increasing
the physical footprint. However, simply continuing to increase physical size is
not a guarantee for further improvements in coherence times as other loss
mechanisms can become dominant. It is therefore important to design qubits that
are sensitive to specific loss mechanisms and perform a systematic study to
optimize qubit coherence times.

\begin{figure}[htbp!]
		 \centering
	\includegraphics[width=0.5\textwidth]{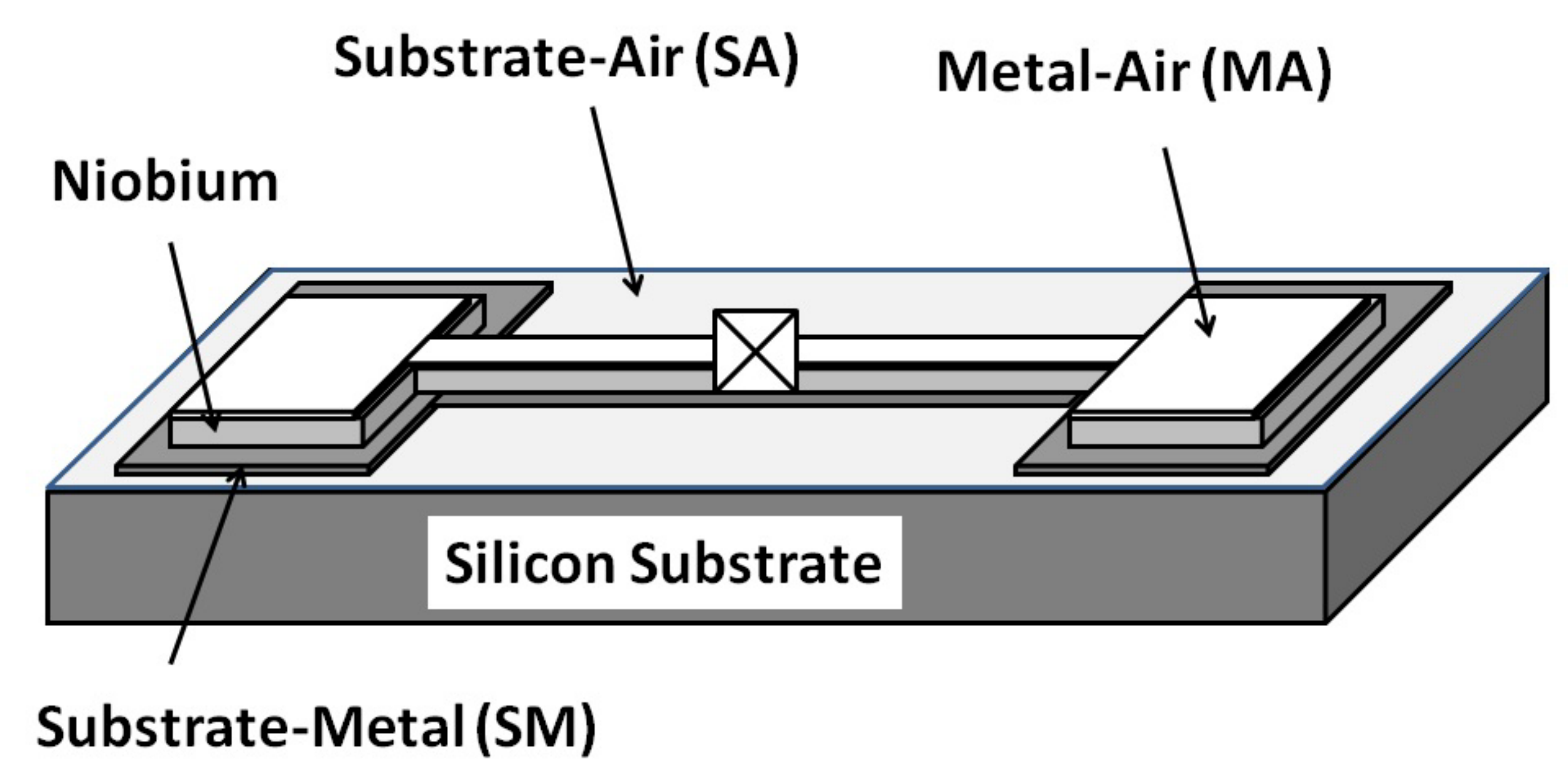}
		 \caption{Schematic of a superconducting qubit fabricated on top of a Silicon substrate. Surface loss can arise due to contributions from the substrate-air (SA) interface, the substrate-metal (SM) interface and the metal-air (MA) interface.}
		 \label{fig:fig1}
\end{figure}

Inspired by similar work for 3D qubits \cite{Dial16}, we here implement a study
for 2D transmon qubits \cite{Koch07} fabricated on high resistivity silicon.
Anticipating energy dissipation from surface loss, we tested several styles of
transmon qubits. From these studies we find that the qubits with physically
small shunting capacitors (i.e. a small physical footprint) show a
characteristic $T_1$ decay which is likely limited by losses associated with one
or more surfaces (see Fig. \ref{fig:fig1}) though present data is not sufficient
to pinpoint which surface dominates. Meanwhile qubits with physically large
footprints show a characteristic $T_1$ that appears to be limited by
contributions of currently unknown origin and saturates near $T_1 \sim
50~\mu\mathrm{s}$, corresponding to a quality factor $Q=T_12 \pi f \sim
1.5~\mathrm{M}$ where $f \sim 4-5~\mathrm{GHz}$ is the qubit frequency.

All presented $T_1$ times are typically averaged over several hours and in most
cases fluctuations on the order of $\pm 20 \%$ have been observed, although
fluctuations by as much as a factor of $2-4$x have been noticed in some rare
cases. We believe these are real $T_1$ fluctuations because the uncertainty from
each fit is typically less than the measured $T_1$ variations with time. Since
$T_1$ itself appears to be fluctuating, the values that are shown should be
interpreted with a sizable error bar. We believe that in the long term it is
useful to introduce metrics to quantify these fluctuations so that experiments or
improvements can be more readily compared. Despite these fluctuations,
meaningful information can be extracted from our experiments described below.

\section{Design and simulation}

A systematic study of coherence times as a function of qubit geometry has been
done for 3D transmon qubits on sapphire \cite{Dial16}, and we adopt similar
strategies here. The experimental goal is to design, fabricate and test transmon
qubits with identical anharmonicities of approximately $E_c \sim 350$ MHz but
substantially different physical sizes for the shunting capacitor to test how
$T_1$ times vary and compare the results to theory.

A total of 5 qubit styles were designed and simulated using Ansys HFSS and Q3D
software, and shown in Figs. \ref{fig:fig4} through \ref{fig:fig8}. We refer to
these different designs as MOD A through MOD E. Participation factors were
determined by calculating the corresponding energy based on the electric fields
present at particular surfaces and interfaces within the build. Hypothetical
layers possessing a dielectric constant of $\epsilon_r=5$ \cite{Dial16} were
placed between the substrate and overlying metallization (SM), the top surface
of the substrate exposed to air (SA) and the metallization surfaces exposed to
air (MA).  Continuous layers of uniform thickness were assumed but because the
thickness of these layers was not known {\em a priori}, the units of the
calculated surface participation factors in Table \ref{table:tab1} are
$\mathrm{m}^{-1}$. For a given layer, multiplying the participation factor by
the thickness and tan($\delta$) gives the reciprocal of the qubit quality factor Q assuming no
other sources of loss.

Because the thicknesses of these contamination layers are presumed to be much
thinner ($<$ 10 nm) than the dimensions of the constitutive elements of the
qubit metallization, the constraints on the domain discretization necessary to
produce a credible numerical solution made conventional simulation impractial. 
Instead we adopted a strategy of calculating the effective electric field
strength at key interfaces by first assuming no contamination layer and
extrapolating the solution to a different dielectric constant than those on
either side of the original interface.  The effective fields were calculated by
matching the appropriate components of the electric fields as prescribed by the
boundary conditions between two dielectric layers:

\begin{eqnarray}
& \vec{E}_{||}^{(1)}  = \vec{E}_{||}^{(2)}   \label{eq:eqn1} \\
& \epsilon_{1} \vec{E}_{\perp}^{(1)}  = \epsilon_{2}\vec{E}_{\perp}^{(2)} \label{eq:eqn2}
\end{eqnarray}
where \(\vec{E}_{||} = -(\vec{E} \times \vec{n}) \times \vec{n}  \) is the component of the field parallel to the interface, \(\vec{E}_{\perp} = \vec{n} (\vec{E} \cdot \vec{n}) \) is that normal to the interface, superscript 1 refers to the hypothetical contamination layer and 2 the actual dielectric material present in the simulation (e.g: silicon for the substrate surfaces or vacuum for the free surfaces).  Combining Equations \ref{eq:eqn1} and \ref{eq:eqn2}, we arrive at the effective electric field in the contamination layer with dielectric constant \(\epsilon_{1} \):

\begin{equation} \label{eq:eqn3}
\vec{E}^{(1)} = \frac{\epsilon_{2}}{\epsilon_{1}} \vec{n} (\vec{E}^{(2)} \cdot \vec{n}) - (\vec{E}^{(2)} \times \vec{n}) \times \vec{n}
\end{equation}
Calculations of the surface participation factors, \( p_{i} / t_{i} \), were
then conducted by projecting a uniform electric field strength throughout the
layer thickness, \( t_{i} \), and integrating over the corresponding surface,
$S$:

\begin{equation} \label{eq:eqn4}
\frac{p_{i}}{t_{i}} \approx \frac{\int_{S} \vec{E}^{(1)} \cdot \epsilon_{1} \vec{E}^{*(1)} dS}{U_{tot}}
\end{equation}
where \(  \vec{E}^{*} \) is the conjugate of \( \vec{E} \) and \( U_{tot} \) is the total energy of the system.

\begin{table}[h]
\resizebox{1\textwidth}{!}{\begin{minipage}{\textwidth}
\begin{tabular}{|c|c|c|c|c|c|}
\hline  & MOD A & MOD B  & MOD C  & MOD D & MOD E \\ 
\hline SM & 3.68x$10^6$ & 1.10x$10^6$ & 3.46x$10^5$ & 2.15x$10^5$ & 1.20x$10^5$ \\ 
\hline SA & 1.24x$10^6$ & 3.32x$10^5$ & 9.10x$10^4$ & 5.04x$10^4$ & 2.38x$10^4$  \\ 
\hline MA & 7.20x$10^4$ & 1.62x$10^4$ & 3.80x$10^3$ & 1.80x$10^3$ & 6.0x$10^2$ \\ 
\hline 
\end{tabular} 
\end{minipage} }
\vspace{0.1in}
\caption{Summary of simulated surface participation to qubit loss for the five different qubit designs with a substrate trench depth of 50 nm. Of particular interest is the significant change in SA participation between MOD A and B. It is also interesting to note that the MA participation is simulated to be very small.}
\label{table:tab1}
\end{table}

\begin{figure}[htbp!] \centering
\includegraphics[width=0.5\textwidth]{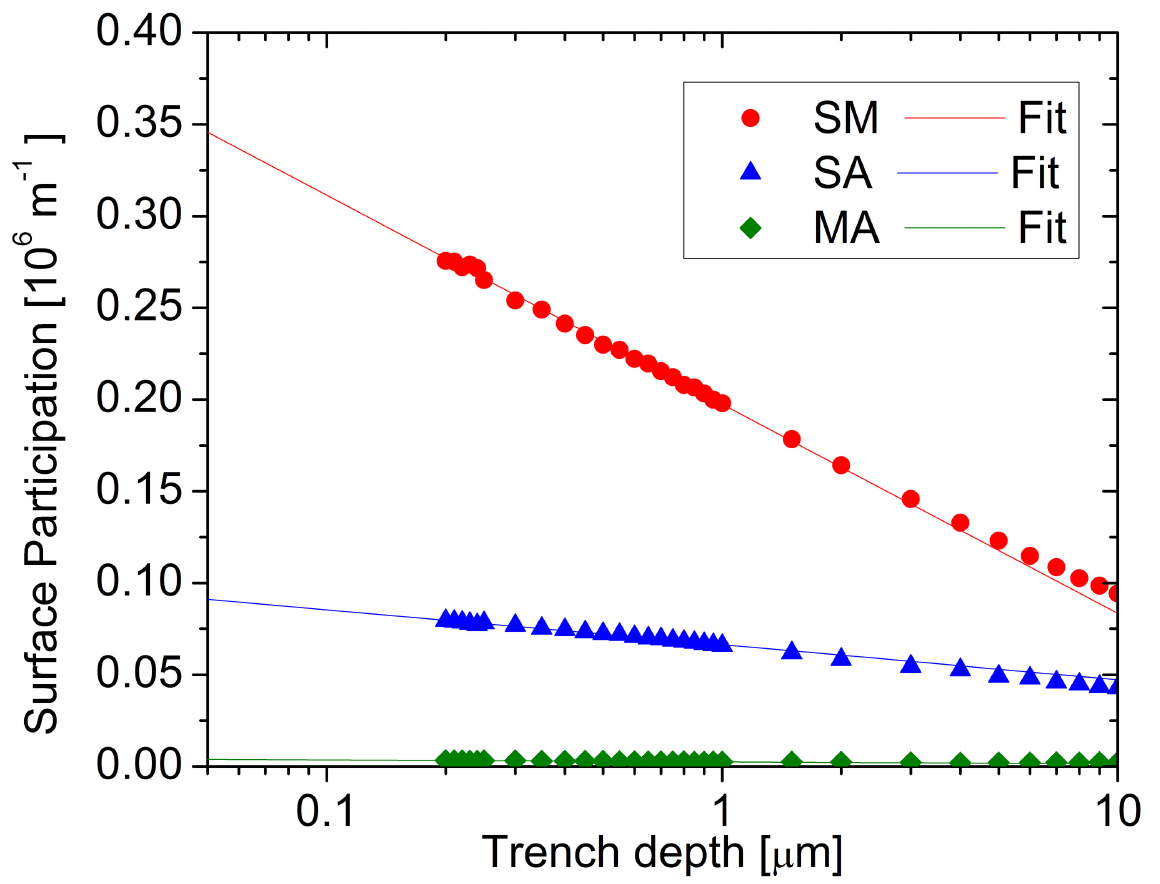} \caption{Calculated
surface participation factors for the substrate-metal (SM), substrate-air (SA)
and metal-air (MA) interfaces as a function of qubit pocket trench depth for the
MOD C qubit design.  The participation factors obey a logarithmic dependence for
small trench depths ($< 1~\mu\mathrm{m}$) which is predicted by the
$r^{-\frac{1}{2}}$ electric field dependence near the metallization edges. Below
trench depths of $300$ nm, surface participation factors were extrapolated using
the logarithmic fits (solid lines) due to simulation convergence issues arising
from the singular behavior of the electric fields.}
\label{fig:figmodcpart}
\end{figure}

Additional complications are that the electric field diverges near corners, and that the etching procedures used in the fabrication of the 2D transmon qubits result in some degree of recess of the substrate surface in the qubit pocket. It is thus important to incorporate the effects of trenching in the surface participation calculations.  HFSS models were generated that included both 200 nm thick Nb metallization and a variation in the pocket trench depth.  Identical qubit junctions, treated as lumped elements, were used for all of the MOD's under investigation.  Because the singular nature of the electric fields near the edges of the Nb metallization can lead to convergence issues in the simulations of trench depths less than $\sim$ 200-300 nm, we restricted the modeling to trenches greater than this thickness and extrapolated participation factors of the various interfaces according to a lograrithmic fit with respect to the trench depth.  Because the electric field strength obeys a $r^{-\frac{1}{2}}$ dependence for a thin metallic sheet \cite{Jackson75} where $r$ is the distance from the sheet edge, the integrand in Equation \ref{eq:eqn4} should scale as $r^{-1}$ for small $r$, leading to the logarithmic dependence in surface participation as confirmed in Figure \ref{fig:figmodcpart}.  Table \ref{table:tab1} contains the extrapolated participation factors to a trench depth of 50 nm for all of the designs, demonstrating that the substrate-metal (SM) interface participation varies by a factor of $30$x over the different designs, the substrate-air (SA) interface participation varies about a factor of $50$x, and while the metal-air (MA) interface participation varies over two orders of magnitude, its participation is substantially less than that of the SM and SA interfaces and is likely not contributing to loss.

Substrate participation was simulated using HFSS in a similar manner as that
used for the surface participation factors. The substrate participation
asymptotes to a constant value in the limit that the trench depth approaches
zero.  In this limit, in which the magnitude of the electric fields present in
the dielectric substrate and in vacuum $(\epsilon_r = 1)$ are equal, the
substrate participation approaches \( \frac{\epsilon_{sub}}{\epsilon_{sub}+1} =
0.92 \) for all of the MOD's. Here and in all simulations we have assumed the
dielectric constant of the substrate as $\epsilon_{sub}=11.45$.

Chip sizes were $4 \times 7$ and $4 \times 8$ $\mathrm{mm}^2$, small enough
that the fundamental chip mode was located above $10$ GHz. Each chip consists of a single
$50$ Ohm feedline, off of which four $\lambda/2$ CPW resonators (also $50$ Ohm)
are each terminated by a transmon qubit of identical design. The resonator
frequencies are staggered at $6.75, 7.0, 7.25, 7.5~\mathrm{GHz}$ with a coupling
quality factor of $10,000 - 40,000$. Given qubit-resonator coupling strengths in the
range of $g=10-50~\mathrm{MHz}$, we calculate the Purcell loss \cite{Houck08} to
be negligible in all our experiments with qubit frequencies targeted in the
$f=4-5~\mathrm{GHz}$ range. The qubit chip is back mounted to a PC board and
placed in a light tight enclosure to minimize exposure to IR radiation
\cite{Barends11,Corcoles11}. Most $T_1$ measurements were done using the high
power readout \cite{Reed10a}. To ensure no residual effects from the readout are
present \cite{Wang14} we varied the repetition time or employed standard low
power QND measurements for many samples (but not all) and observed no noticeable
differences. The input attenuation and filtering to the sample varied slightly
between experiments to test thermal dephasing \cite{Rigetti12}, but these
variations were found to have little impact, if any, on $T_1$ times. However,
because $T_2$ times are typically near $T_1$ there is still room for improvement
for future experiments. The output lines included PAMTECH $3-12$ GHz isolators
(model CWJ0312KI) and a 4K HEMT amplifier by CalTech.

\section{Fabrication,testing and results}

The qubits were fabricated on high resistivity ($\rho \sim 1k \Omega$cm) silicon
using a standard niobium (Nb) recipe with a hydrogen fluoride (HF) clean just
prior to the deposition of the sputtered Nb. The junctions are fabricated using
a cold develop and acetone lift off at room temperature. Each wafer received a
Huang clean to remove organic and inorganic contaminants from the wafer surface.
 Within 30 minutes prior to the 200 nm Nb deposition each wafer received a 2
min. 100:1 HF clean to remove the surface SiOx. After the HF treatment each
wafer received a 5 min water rinse.  After the water rinse and dry there is a
thin SiOx layer on the surface. Next, a chlorine based RIE etch was performed
which slightly over etches the silicon, etching approximately $50-100$ nm into
the silicon. The aluminum - aluminum oxide - aluminum Josepshon junction is
fabricated using a Dolan bridge \cite{Dolan77} employing PMMA/MMA processing
steps. The aluminum makes contact to the niobium after applying a gentle low
power ion mill clean prior to the aluminum deposition. We have tested additional
qubits (data not shown) with slight modifications to the pre-cleans prior to the
niobium deposition but found no deviations from the data reported here, and
similarly a junction lift-off in Microposit Remover 1165 at $80^{\circ}$C was
also found not to affect coherence times.

\begin{figure}[htbp!]
		 \centering
	\includegraphics[width=0.5\textwidth]{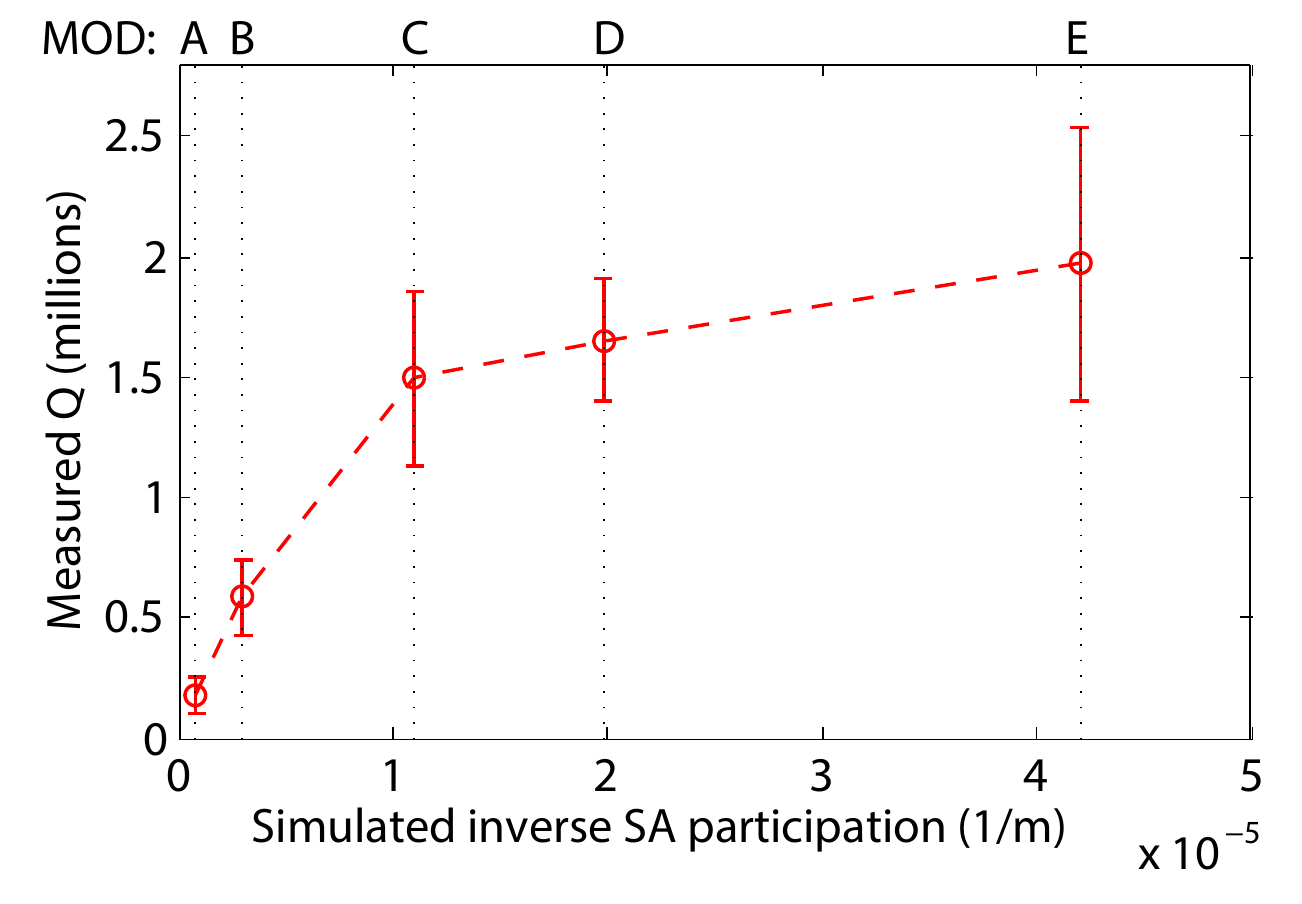}
		 \caption{Qubit quality factor as a function of the inverse SA surface participation. Markers indicate the average, and  error bars indicate $\pm$ the standard deviation divided by the square root of the number of measured qubits. The data indicates that low MOD qubits are consistent with surface loss due to linear scaling with the simulation. Coherence of high MOD qubits appears to be much less sensitive to surface loss, seemingly limited by another loss mechanism.}
		 \label{fig:fig3}
\end{figure}

The measured $T_1$ times are translated into quality factors in order to remove
effects from frequency variations of the qubits; within the range of qubit
frequencies measured, no frequency dependence of the quality factor was found.
The results are summarized in Fig. \ref{fig:fig3}, where the measured quality
factors are plotted against the inverse of simulated SA surface loss
participation values. If coherence times were limited by SA loss we would expect
Q to linearly scale with the inverse of the participation values. Each data
point represents the mean quality factor of eight qubits, and the error bar is
the standard deviation divided by the square root of the number of measured
qubits. Each $T_1$ is obtained by averaging typically on the order of a few
hours to account for known fluctuations. We observe a dramatic improvement in Q
between MOD A and MOD B qubits, consistent with predictions based on surface
loss. When plotted against SM or MA inverse surface participation values (not
shown) the results look very similar, which leads us to the conclusion that at
present the data is not sufficient to ascertain which of the interfaces
dominates. High MOD qubits appear to saturate near $Q \sim 1.5-2~\mathrm{M}$
indicating a loss contribution of unknown origin, which could potentially
include substrate loss, loss due to quasi-particles \cite{Wang14}, residual
coupling to resistive metallic components housing the qubits or other
mechanisms. Nonetheless, it is possible to place an upper bound on the bulk loss
of the silicon of $\mathrm{tan}(\delta)\sim 5\times 10^{-7}$ though we believe
the actual number is likely less.

\section{Conclusion}

In conclusion, we present a systematic study of quality factors in qubits with
widely varying degrees of surface participation. Results indicate that depending
on the design, qubits can be either limited by surface loss (low-MOD qubits) or
an additional loss mechanism (high-MOD qubits) that is currently not well
understood. The experimental results are not sufficient for us to make any
claims on which surface dominates surface loss.  We believe the methods applied
here should allow for optimal design of qubits for long coherence to enable
relevant multi-qubit implementations. The results also motivate the development
of cleaning methods to potentially enable smaller footprint qubits with good
coherence in the future, as well as the development of additional strategies to
understand new loss mechanisms for high MOD qubits. We would like to acknowledge
useful discussions and contributions from Mary Beth Rothwell, George Keefe and
Cyril Cabral.

\section{Pictures of the qubit designs}
Below are figures showing each of the five MOD designs in detail. Note that all
qubits are placed inside a grounding box of dimension $650\times
650~\mu\mathrm{m}^2$. This box is only fully visible for MODs D and E. MODs A
and B were also tested with a larger coupling capacitor to increase the
signal-to-noise ratio. The data indicates no appreciable effect on $T_1$, which
is not surprising given that the calculated Purcell limit is much larger than
the observed coherence times.

\begin{figure}[htbp!] \centering
\includegraphics[width=0.5\textwidth]{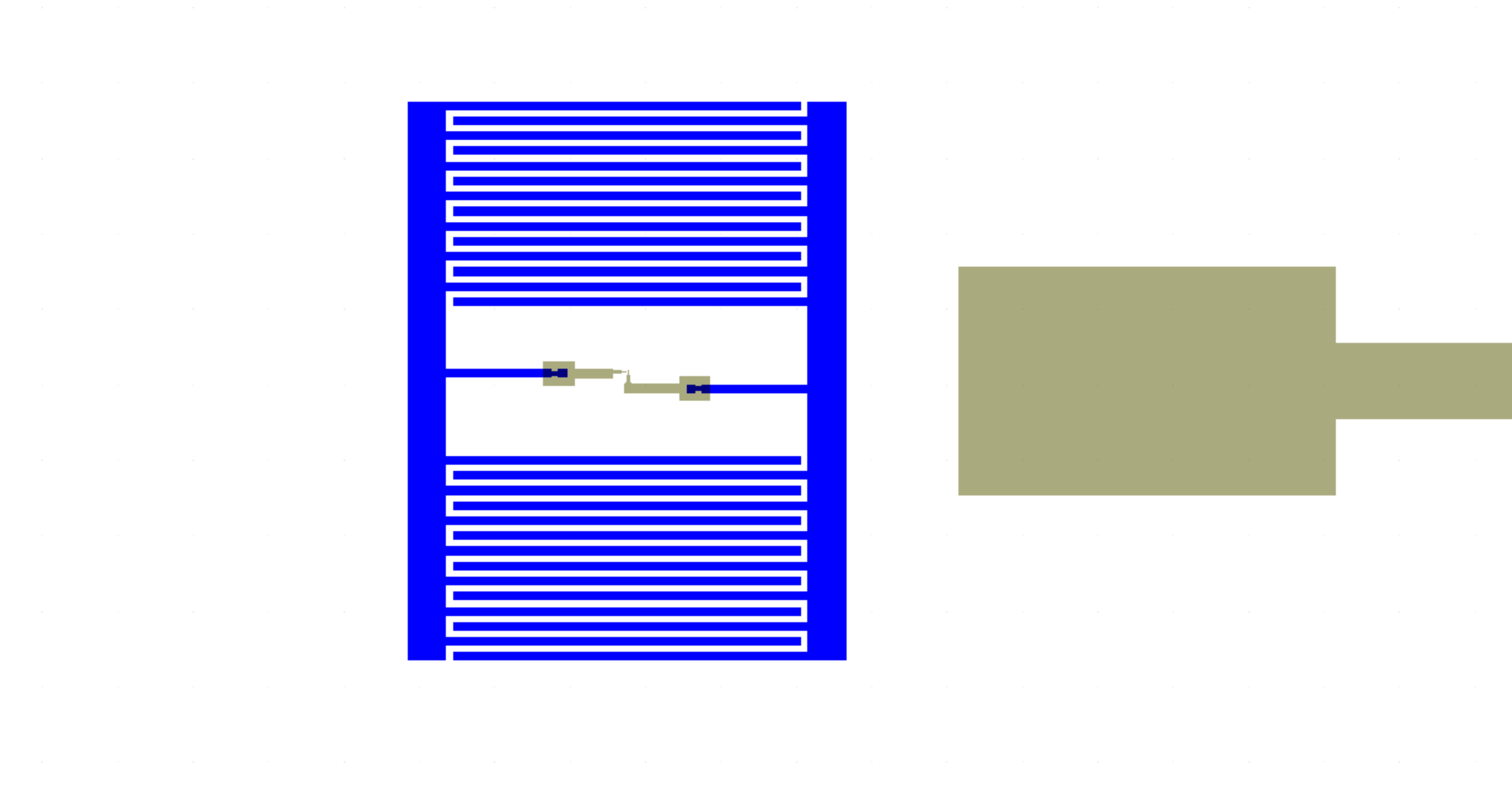}
		 \caption{Schematic of the MOD A design featuring an interdigitated capacitor with a finger linewidth and finger-to-finger spacing of $1$ $\mu$m. The rectangle to the right is the coupling capacitor to the CPW readout resonator of sufficient size such that $g=8$ MHz.}
		 \label{fig:fig4}
\end{figure}

\begin{figure}[htbp!]
		 \centering
	\includegraphics[width=0.5\textwidth]{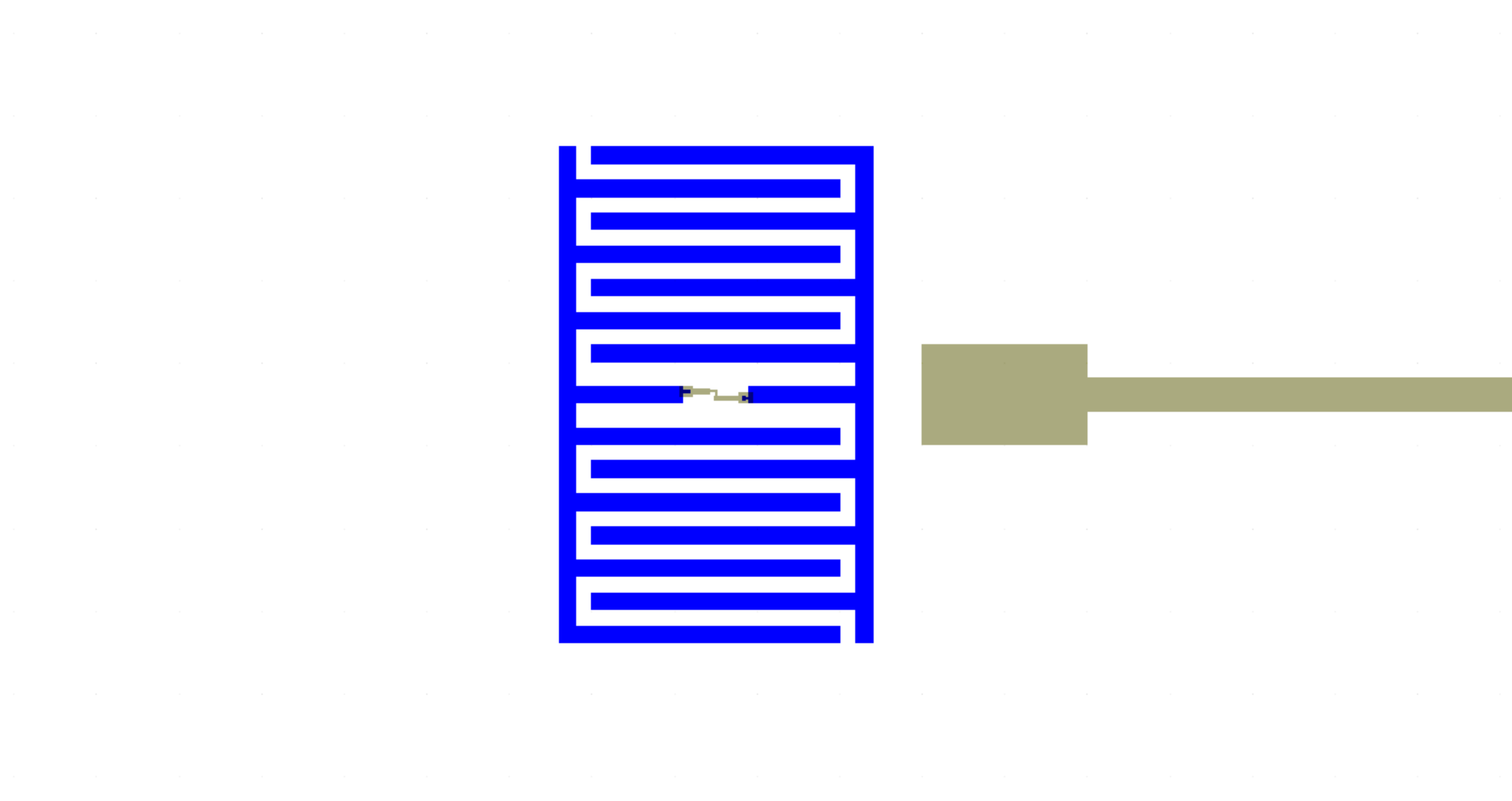}
		 \caption{Schematic of the MOD B design featuring an interdigitated capacitor with a finger linewidth and finger-to-finger spacing of $5$ $\mu$m. The rectangle to the right is the coupling capacitor to the CPW readout resonator of sufficient size such that $g=20$ MHz.}
		 \label{fig:fig5}
\end{figure}

\begin{figure}[htbp!]
		 \centering
	\includegraphics[width=0.5\textwidth]{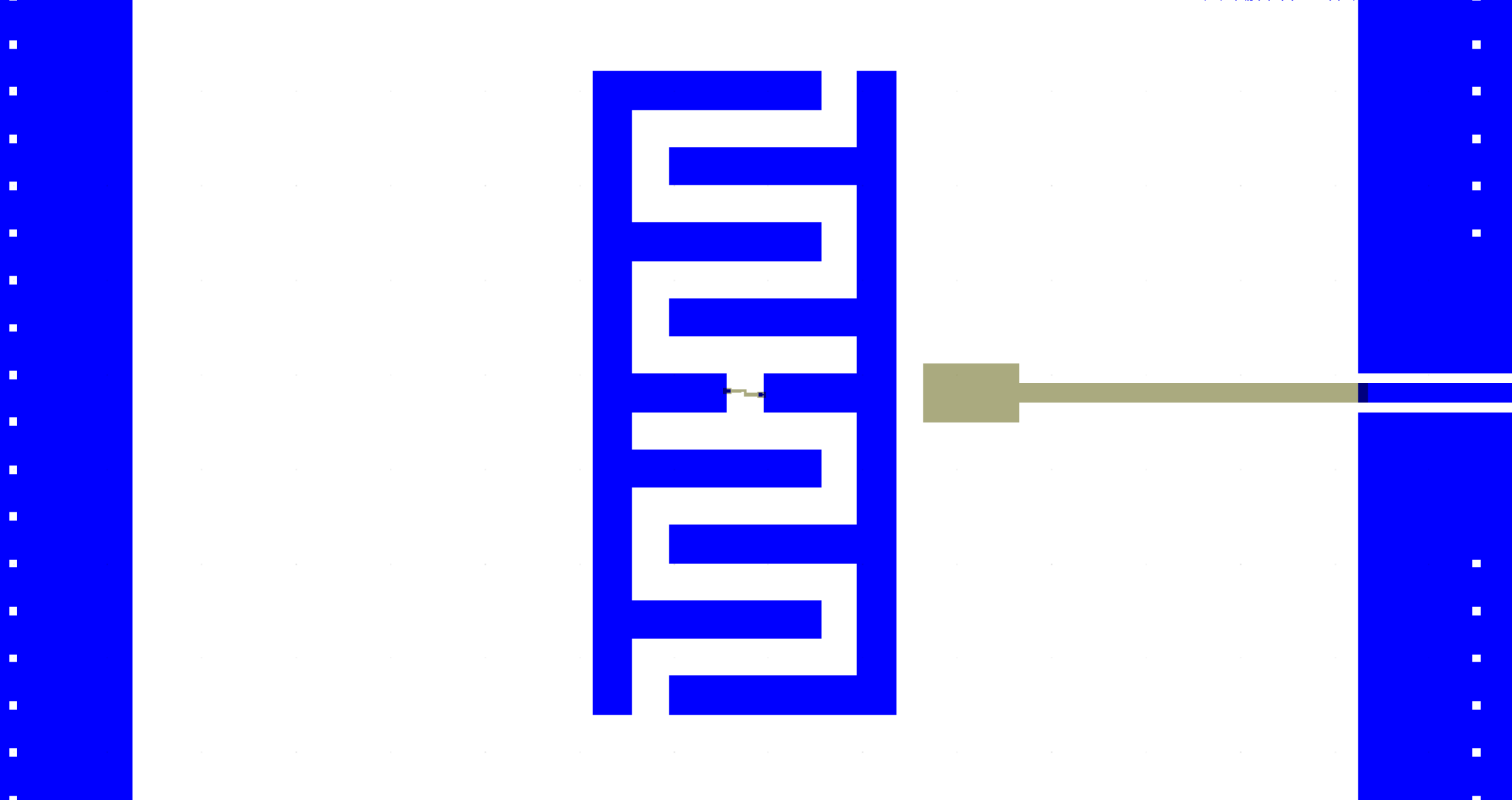}
		 \caption{Schematic of the MOD C design featuring an interdigitated capacitor with a finger linewidth and finger-to-finger spacing of $20$ $\mu$m. The rectangle to the right is the coupling capacitor to the CPW readout resonator of sufficient size such that $g=45$ MHz.}
		 \label{fig:fig6}
\end{figure}

\begin{figure}[htbp!]
		 \centering
	\includegraphics[width=0.5\textwidth]{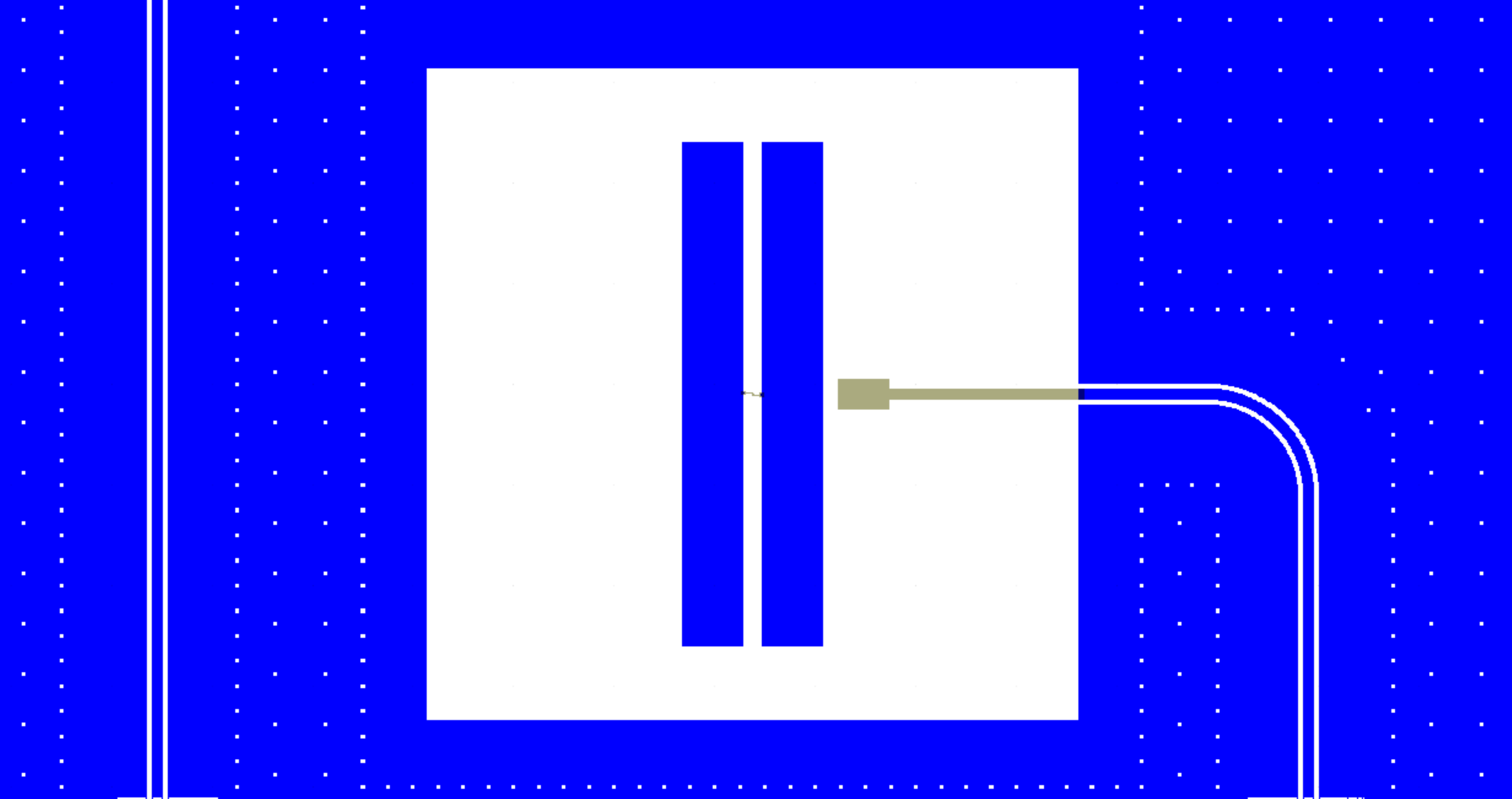}
		 \caption{Schematic of the MOD D design featuring two capacitor pads of
		 dimension $500\times 60~\mu\mathrm{m}^2$ separated by $20~\mu\mathrm{m}$. The
		 rectangle to the right is the coupling capacitor to the CPW readout resonator of sufficient size such that $g=52$ MHz.}
		 \label{fig:fig7}
\end{figure}

\begin{figure}[htbp!]
		 \centering
	\includegraphics[width=0.5\textwidth]{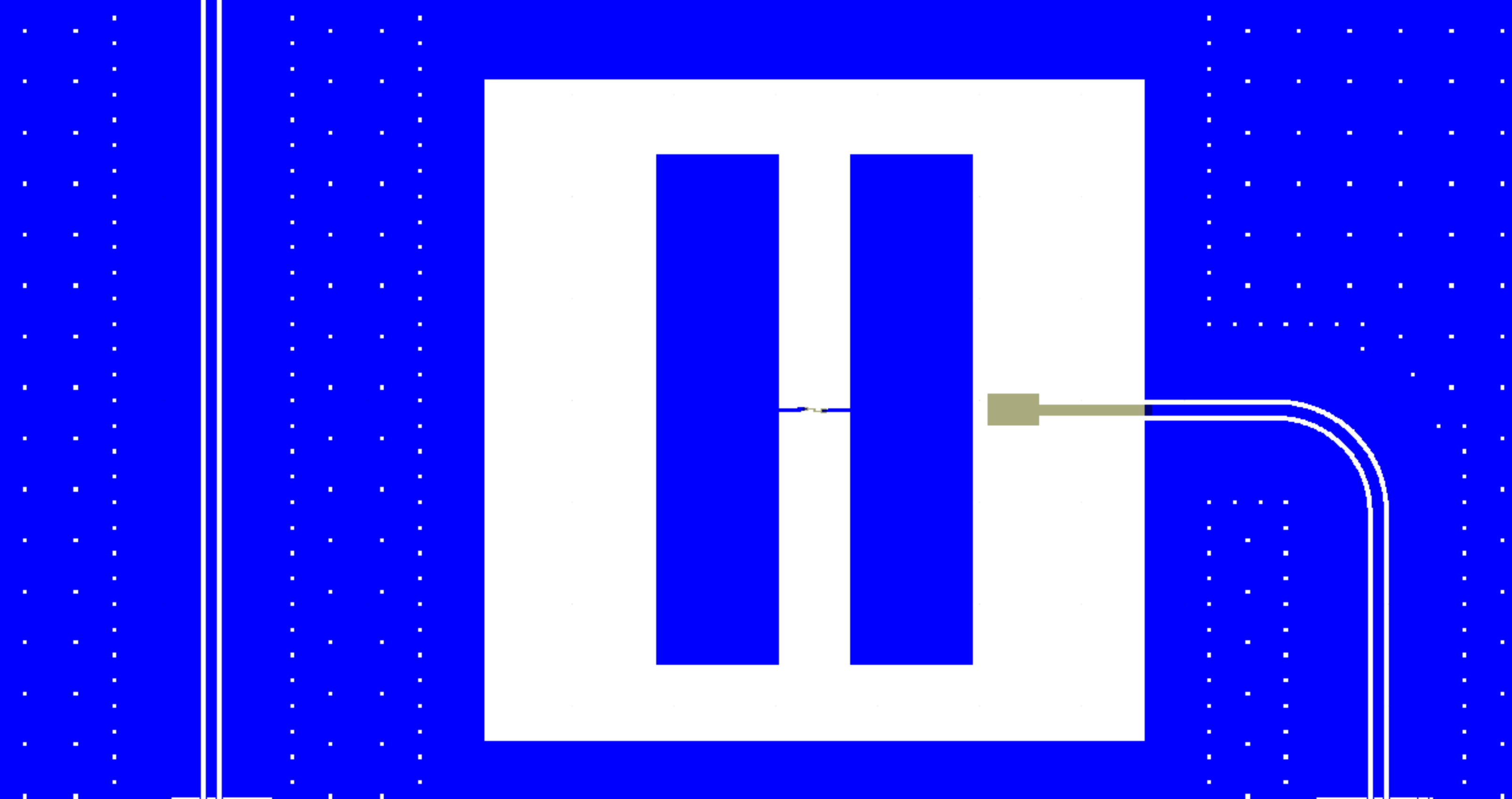}
		 \caption{Schematic of the MOD E design featuring two capacitor pads of
		 dimension $500\times 120~\mu\mathrm{m}^2$ separated by $70~\mu\mathrm{m}$.
		 The rectangle to the right is the coupling capacitor to the CPW readout resonator of sufficient size such that $g=53$ MHz.}
		 \label{fig:fig8}
\end{figure}

\bibliography{bibmaster_cem}

\end{document}